\documentclass[conference]{IEEEtran}
\IEEEoverridecommandlockouts
\usepackage{cite}
\usepackage{amsmath,amssymb,amsfonts}
\usepackage{algorithmic}
\usepackage{graphicx}
\usepackage{textcomp}
\usepackage{xcolor}
\def\BibTeX{{\rm B\kern-.05em{\sc i\kern-.025em b}\kern-.08em
    T\kern-.1667em\lower.7ex\hbox{E}\kern-.125emX}}
   
 \usepackage{hyperref}

\usepackage{multirow}
\usepackage{color}
\usepackage[section]{placeins}

\usepackage[greek,english]{babel}
\languageattribute{greek}{polutoniko}
\usepackage[iso-8859-7]{inputenc}

\def\dalemb#1#2{{\vbox{\hrule height.#2pt
  \hbox{\vrule width.#2pt height#1pt \kern#1pt \vrule width.#2pt}
    \hrule height.#2pt}}}

\def\ba{\begin{eqnarray}}
\def\ea{\end{eqnarray}}
\def\be{\begin{equation}}
\def\ee{\end{equation}}

\def\gtorder{\mathrel{\raise.3ex\hbox{$>$}\mkern-14mu
             \lower0.6ex\hbox{$\sim$}}}
\def\ltorder{\mathrel{\raise.3ex\hbox{$<$}\mkern-14mu
             \lower0.6ex\hbox{$\sim$}}}

\begin{document}

\title{A deep--learning classifier for cardiac arrhythmias
}

\author{\IEEEauthorblockN{Carla Sofia Carvalho}
\IEEEauthorblockA{
\textit{Hitachi Vantara}\\
Lisbon, Portugal \\
carla.carvalho@hitachivantara.com}
}

\maketitle

\begin{abstract} 
We report on a method that classifies heart beats according to a set of 13 classes, including cardiac arrhythmias. 
The method localises the QRS peak complex to define each heart beat and uses a neural network to infer the patterns characteristic of each heart beat class. The best performing neural network contains six one--dimensional convolutional layers and four dense layers, with the kernel sizes being multiples of the characteristic scale of the problem, thus resulting a computationally fast and physically motivated neural network. 
For the same number of heart beat classes, our method yields better results with a considerably smaller neural network than previously published methods, which renders our method competitive for deployment in an internet--of--things solution. 
\end{abstract}




\section{Introduction}

An industry domain that inherently produces data is the health domain, in particular the subdomain related to the  monitoring of patients. 
Since cardiovascular diseases are the first cause of death worldwide 
\footnote{\url{https://www.who.int/gho/mortality_burden_disease/causes_death/top_10/en/}}, 
a common monitoring is that of the heart as a way to identify or prevent heart dysfunctions. 

Heart dysfunctions are related to anomalies in the heart's electrical activity, including cardiac arrhythmias,  
and can be diagnosed in electrocardiograms (ECG), 
produced in real time by  
portable devices. These records show the heart's beating patterns in time as a result of differences in the electrical potential in the heart. 

The interest thus lies in producing a 
physically motivated method that classifies heart beats from medically annotated ECG records in a fast and robust way, so that it can be deployed to generate alerts. 
Our suggested method encompasses the processing of ECG records, based on the location of characteristic features, 
and a classification model, based on a deep neural network, with the medical annotations providing the labels. 

The advantage of neural networks over a heuristic model is that  
they search for the optimal combination of weights over different layers in sequence, which add non--linearities and can reproduce different functional forms. 
Neural networks have been used in the past to classify cardiac arrhythmias, using different number of heart beat types (e.g. Refs.~\cite{acharya_2017, he_2018, jun_2018}) and adopting complex architectures with difficult interpretability (e.g. Ref.~\cite{rajpurkar_2017}).
In this paper, we test different architectures from previously published work \cite{acharya_2017, 
gupta_2018} and create new neural networks based on the scale of the problem, with view towards increasing the performance while simultaneously keeping the neural networks short and fast. 



\begin{figure}
\centerline{
\includegraphics[width=7.cm]{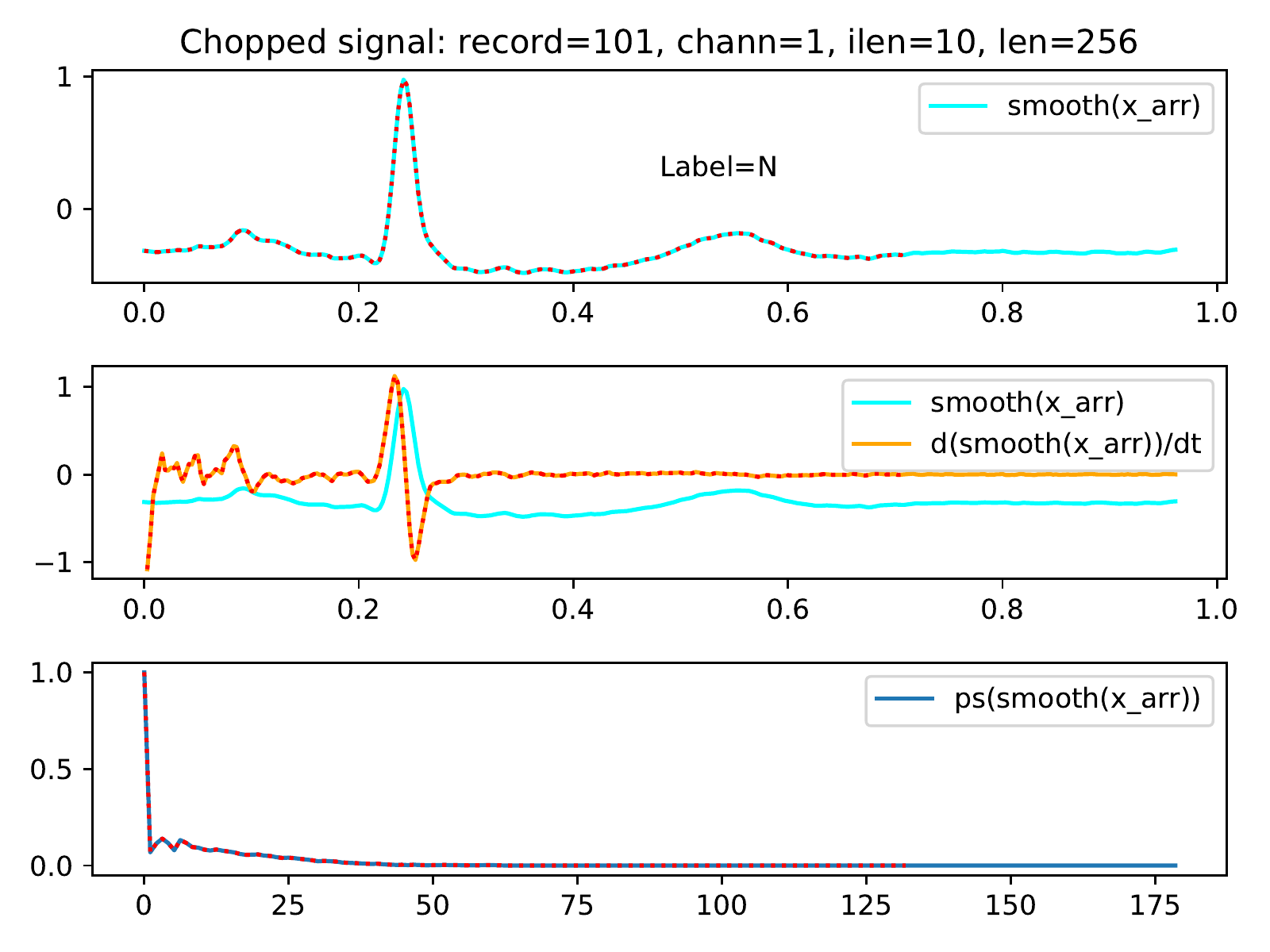}}
\centerline{
\includegraphics[width=7cm]{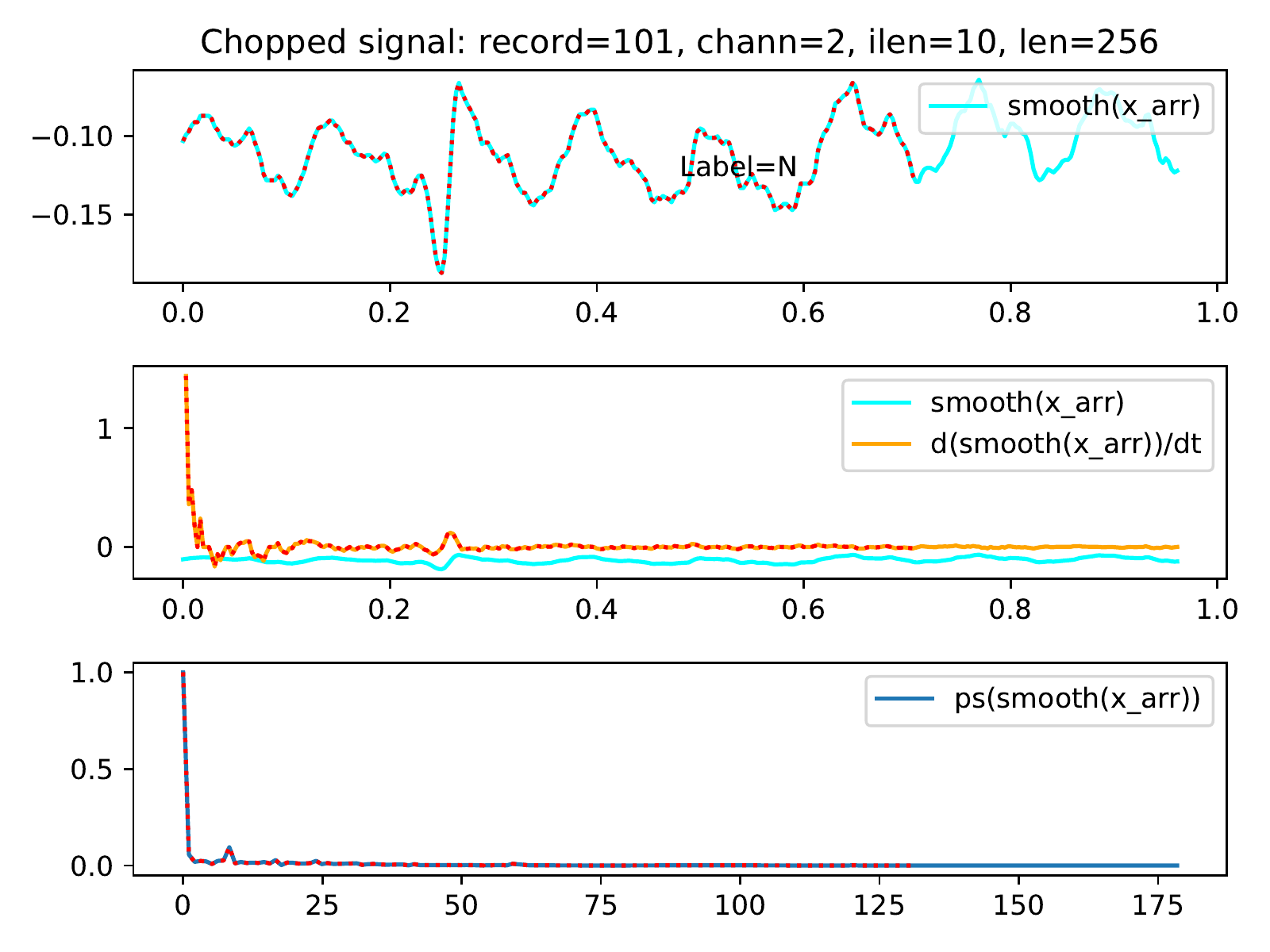}}
\vspace{-0.2cm}
\caption{\baselineskip=0.4cm{
{\bf Signals in record 101.}  
Chopped smoothed signals and their subsequent discrete first derivative and amplitude spectrum. Top three panels: Channel 1. Bottom three panels: Channel 2.}}
\label{fig:chop_signal}
\vspace{-0.2cm}
\end{figure}

\section{Data processing}
\label{sec:data_proc}

\subsection{Data digitisation}

We use publicly available data from the MIT--BIH Arrhythmia Database Directory, comprising 48 records\footnote{\url{https://physionet.org/physiobank/database/html/mitdbdir/mitdbdir.htm}}. These records contain signals from two ECG channels (an upper signal and a lower signal) sampled at a frequency ${\tt f_{s}}=1/{\tt dt}=360~\rm{Hz}$ for ${\tt N} \times{\tt dt}=30~\rm{min}.$  
These records also contain annotations by two cardiologists.  
Some records contain paced beats driven by a pacemaker or artifacts. We choose to use all 48 records, since the paced beats can work as an additional heart beat type 
and the artifacts can work as a noise component.  

We also use the WFDB software package to read and process the 
file format\footnote{\url{https://wfdb.readthedocs.io/en/latest/}} that the records are encoded in.

\subsection{Heart beat locations}

The signals consist of readings of the heart potential in time. The heart potential contains characteristic peaks, namely the P peak, the QRS peak complex and the T peak, which correspond to the polarisation/depolarisation heart cycle.  
The first step consists in identifying the QRS peak complexes in time, which are usually more prominent in the upper signal. 

The annotations are located at the QRS peak complex and provide the labels to the heart beats. 
Hence the next step consists in chopping the signals about each QRS peak complex so that each fraction contains an individual heart beat  
(Fig.~\ref{fig:chop_signal}). Each record has a characteristic beat length between consecutive QRS peak complexes. We choose the median characteristic beat length (${\tt len}=256$) so that the resulting chopped signals 
can be concatenated into a matrix 
${\tt x_{ijk}},$ 
where 
${\tt i}\in \{1,..., {\tt n\_sample}\}$ is the number of resulting chopped signals, 
${\tt j}\in\{1,..., {\tt len}\}$ is the length in time of each chopped signal and 
${\tt k}\in \{{\tt sign1}, {\tt sign2}\}$ indicates the signal. 

\subsection{Heart beat annotations}

The possible values of the annotations define the set of heart beat classes.
The records in the MIT--BIH Arrhythmia Database Directory follow the annotation system such that beat annotations take the values 
$\{{\tt N, L, R, B, a, J, A, S, j, e, n, V, r, E, F, /, f, Q, ?}\}$
and non--beat annotations take all the other possible values \footnote{\url{https://archive.physionet.org/physiobank/annotations.shtml}}. 
 
\begin{figure}
\centerline{
\includegraphics[width=6.5cm]{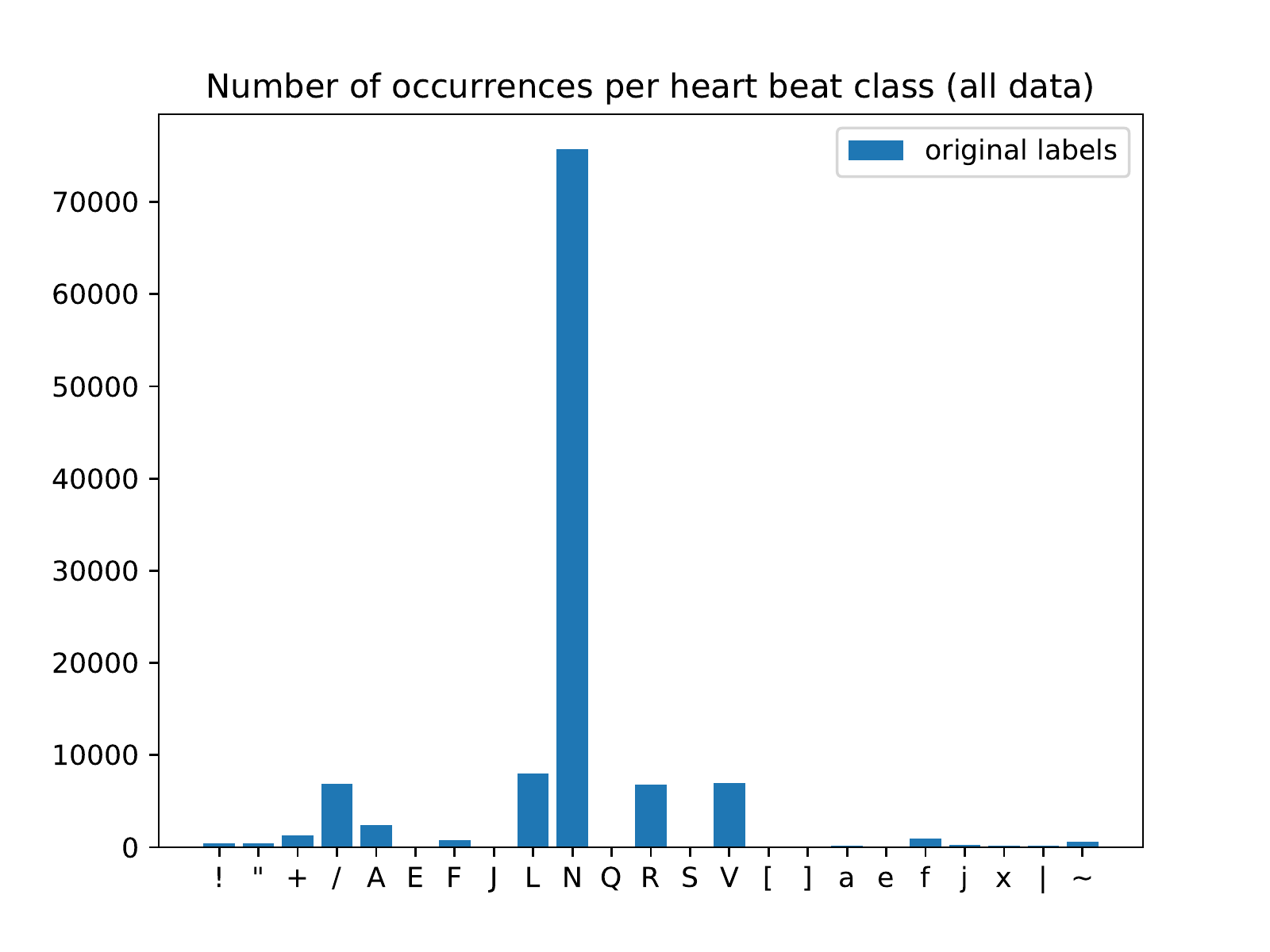}}
\vspace{-0.2cm}
\centerline{
\includegraphics[width=6.5cm]{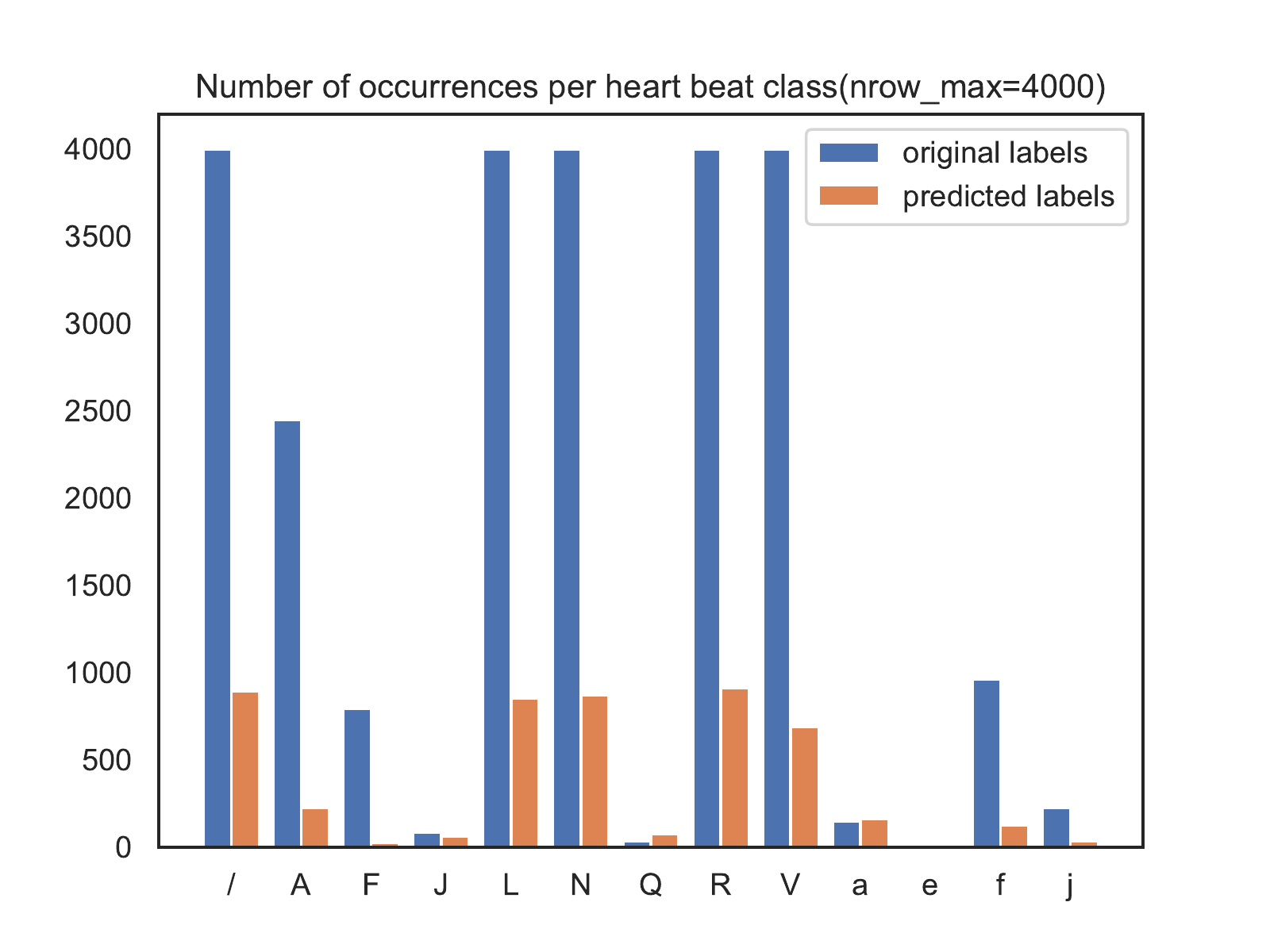}}
\vspace{-0.2cm}
\caption{\baselineskip=0.4cm{
{\bf Distribution of the heart beat classes.} 
Top panel: 
Distribution of the original labels.
Bottom panel: 
Distribution of both the original and the predicted labels (best method),
}}
\label{fig:hist_class}
\vspace{-0.2cm}
\end{figure}

We produce the distribution of heart beat classes across the records (Fig.~\ref{fig:hist_class}, top panel).
We observe that the classes $\{{\tt B, n, r, ?}\}$ are not represented in this data set. We also observe that 
the beat classes are not all equally represented. 
Hence we set an upper bound to the number of occurrences per heart beat class (here ${\tt n\_row\_max}=4000$), so that all beats from under--represented classes are included but only a fraction of the beats from  over--represented classes is included.  
We also observe that the classes $\{{\tt S, E}\}$ contain less than six elements, which is the minimum number of elements required to balance the representation of a given class (Sec.\ref{sec:data_eng}).  
Note that some values do not correspond to heart beat classes, e.g. $\{{\tt Q}\}$ corresponds to unclassifiable beats and $\{{\tt /, f}\}$ corresponds to paced beats; we choose to keep them to add robustness to the model. 
Hence the heart beat classes 
that can be classified are reduced to $\{{\tt N, L, R, a, J, A, j, e, V, F,  Q, /, f}\},$ totalling 13 classes.

We encode the annotation values into binary vectors with length equal to the number of classes, resulting in a matrix ${\tt y_{ic}},$ where 
${\tt i}\in \{1,..., {\tt n\_sample}\}$ is the number of heart beats 
and ${\tt c} \in \{1,..., {\tt n\_class}\}$ is the number of heart beat classes.  


\section{Data engineering}
\label{sec:data_eng}

\subsection{Data resampling}

Since the heart beat classes are not all equally represented, 
we re--sample the data by 
generating synthetic beats belonging to the under--represented classes,  
thus producing a new data set with balanced classes. We use the Synthetic Minority Oversampling Technique ({\sc SMOTE}) as implemented in the imbalanced--learn package
{\footnote{\url https://imbalanced-learn.readthedocs.io/en/stable/index.html}}.
When we  
resample data, the training set undergoes resampling, whereas the test set does not, so that the performance metrics refer to the original class distribution. 

\subsection{Generation of new variables}

From the original signals, we can generate new variables that encode potentially useful information, e.g. the first discrete derivative of the smoothed signals ${\tt dx_{ijk}}$ 
and the Fourier transform of the chopped smoothed signals ${\tt X_{ilk}}.$ 
The Fourier transform of a signal contains both positive--frequency and negative--frequency components; hence, instead of ${\tt X_{ilk}},$ we use the amplitude spectrum $\vert{\tt X_{ilk}}\vert=\sqrt{{\tt X_{ilk}}{\tt X_{ilk}}^{\ast}}.$ 
The result of the concatenation of the original variables with the generated variables is the data matrix
${\tt \tilde x_{ijk}}=\{{\tt x_{ijk}}, {\tt dx_{ijk}}, \vert{\tt X_{ijk}}\vert\}$ such that, for ${\tt n\_chann}=2,$ the data matrix ${\tt \tilde x_{ijk}}$ consists of $3\times {\tt n\_chann}=6$ variables. 

\subsection{Selection of variables}

\begin{figure}[t]
\centerline{
\includegraphics[width=7.cm]{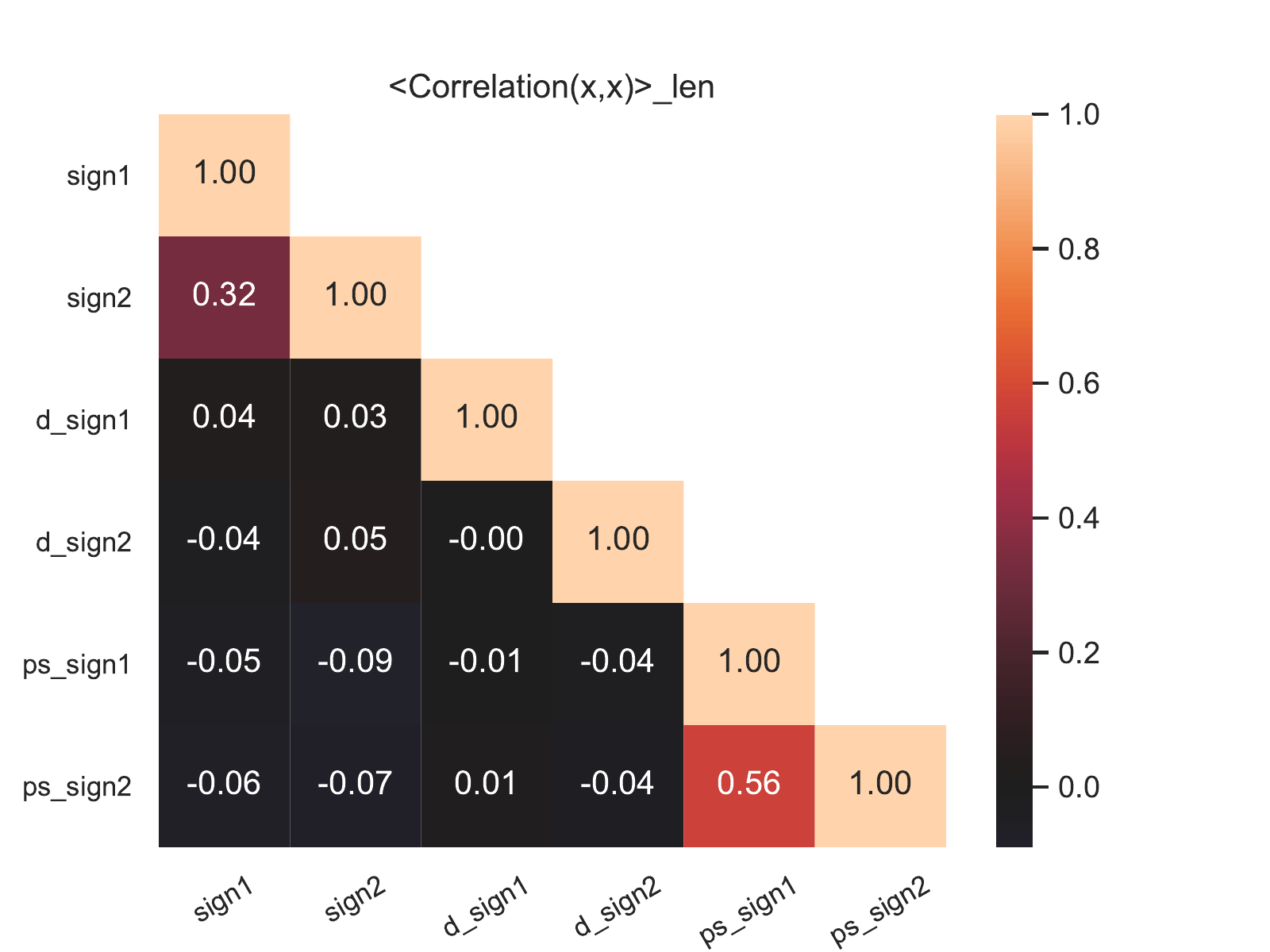}}
\centerline{
\includegraphics[width=7.cm]{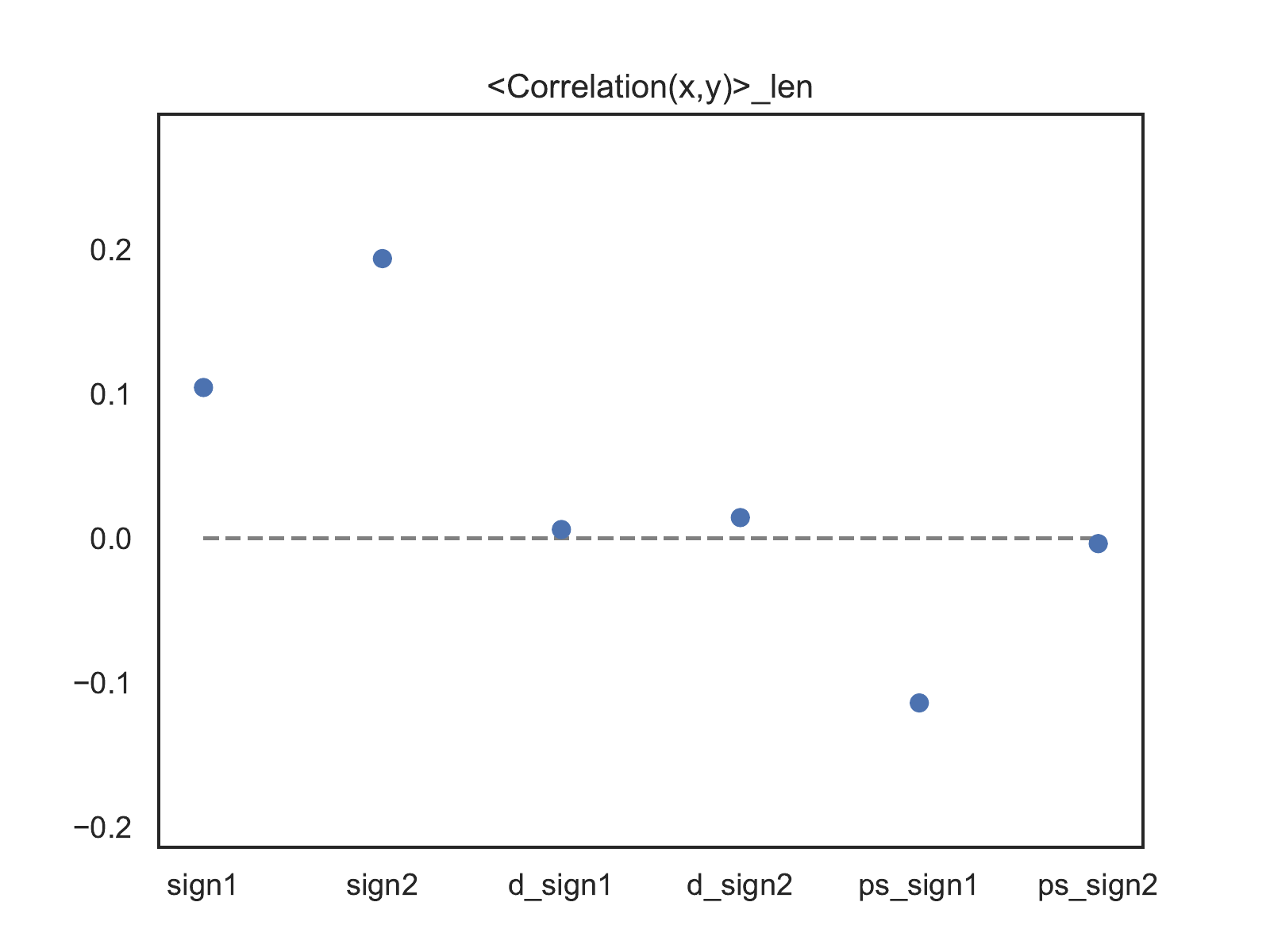}}
\vspace{-0.2cm}
\caption{\baselineskip=0.4cm{
{\bf Correlation matrices.} 
Top panel: Correlation matrix between each pair of variables. 
Bottom panel: Correlation between each variable and the heart beat annotations.}}
\label{fig:corr}
\vspace{-0.2cm}
\end{figure}

We compute the correlation between each pair of variables indexed $\{\tt k_1,k_2\}$ with values $\{\tt \tilde x_{ijk_1} , \tilde x_{ijk_2}\},$ 
which we denote by $\text{Corr}_{\tt k_1k_2}$ (Fig.~\ref{fig:corr}, top panel). 
We also compute the correlation of each variable ${\tt k_1}$ with the beat annotations ${\tt y_{ic}},$  which we denote by $\text{Corr}_{\tt k_1y}$ (Fig.~\ref{fig:corr}, bottom panel). 

By setting an upper bound (e.g. ${\tt corr\_max}=0.9$), 
we use the correlation between each pair of variables as a measure of redundancy. Variables $\{\tt k_1,k_2\}$ such that $\vert \text{Corr}_{\tt k_1k_2}\vert >{\tt corr\_max}$ are classified as redundant, hence one of them can be removed without loss of information. 
From Fig.~\ref{fig:corr} top panel, 
no removal is justified on the basis of redundancy. 

By setting a lower bound (e.g. ${\tt corr\_min}=0.1$),
we use the correlation between each variable with the beat annotations as a measure of relevance of that variable in predicting annotations. A variable ${\tt k_1}$ such that  $\vert \text{Corr}_{\tt k_1y}\vert <{\tt corr\_min}$ is classified as irrelevant, hence it can be removed without loss of information.
From Fig.~\ref{fig:corr} bottom panel, the variables $\{\tt dx_{ij,k=sign1}, dx_{ij,k=sign2}, \vert X_{ij,k=sign2}\vert\}$ can be removed on the basis of relevance. 


\section{Classification models}
\label{sec:class}

\subsection{Neural networks}
Since the heart beats in the data are labelled, we look for a classification model to infer the patterns common to heart beats in the same class. Given the nature of the data, 
neural networks (NN) either of the recurrent type (RNN) or the convolutional type (CNN) will be adequate models.

An RNN regards each heart beat as a sequence of data points in time and combines the value at the previous instant with a transformation of earlier values.  
A variation of RNN is the large--short--term--memory (LSTM) NN. 
Each heart beat 
must be further divided into sublengths of the original beat length so that the different sublengths are regarded as a sequence of data points. 
Since each heart beat has size ${\tt len}$ and we are looking for three peak--like structures, then the characteristic sublength will be ${\tt len}/4 = 64.$ For sublengths, we use fractions of the characteristic sublength, 
in particular 
${\tt sublen} \in \{1/8,1/4,1/2,1\}\times {\tt len}=\{32,64,128,256\}.$ 

A CNN regards each heart beat as a one--dimensional image and operates one--dimensional convolutions (Conv1D) over the
kernel.   
Since the characteristic sublength is ${\tt len}/4,$ then the largest scale will be the length corresponding to the Nyquist frequency, 
hence ${\tt len}/4/2 = 32.$ For kernel sizes, we use powers of two between 4 and 32; 
for stride step, we use ${\tt stride}=1.$ 

For optimiser, we use ADAM (with the default values ${\tt lr}=0.001, {\tt beta\_1}=0.9, {\tt beta\_2}=0.999, {\tt decay}=0$); for loss function, we use categorical cross entropy; for number of epochs, we use ${\tt n\_epochs}=25;$ for batch size, we use ${\tt batch\_size}=64;$ or activation function, we use the rectified linear unit (ReLU).

\begin{table*}[t]
\begin{center}
\caption{\baselineskip=0.4cm{
{\bf Selected architectures for testing.}
Column 1: Identification of the architecture. 
Columns 2-4: Identification of the performance metrics, where Accuracy is the resulting accuracy from the combined classification over all ${\tt nk^{\prime}}$ sets, and 
$\left<\text{Precision}\right>$ and $\left<\text{Recall}\right>$ 
are respectively the mean precision and mean recall over all classes. Column 5: Identification of the efficiency metric, where Run time is the running time in minutes.
}}
\vspace{-0.2cm}
\begin{tabular}{c|cccc}
Architecture & Accuracy & $\left<\text{Precision}\right>$ &$\left<\text{Recall}\right>$& Run time \\
&&&&(min)\\
\hline
${\tt LSTM}$(${\tt sublen}=256$)
			& 0.712 
			& 0.479 & 0.580 & 436\\
${\tt LSTM}$(${\tt sublen}=32$)
			& 0.699
			& 0.460 & 0.556 & 445\\
\hline
${\tt Conv}$
			& 0.781 
			& 0.586 & 0.567 & 101\\
\hline
${\tt Conv}+{\tt LSTM}$
(${\tt sublen}=256$)
			& 0.770 
			& 0.572 & 0.569 & 215\\
${\tt Conv}+{\tt LSTM}$
(${\tt sublen}=32$)
			& 0.761 
			& 0.513 & 0.605 & 143\\
\hline
${\tt ConvLSTM}$
(${\tt sublen}=256$)
			& 0.759 
			& 0.564 & 0.549 & 536\\
${\tt ConvLSTM}$
(${\tt sublen}=32$)
			& 0.759
			& 0.570 & 0.584 & 391\\
\end{tabular}%
\label{table:test_arch}
\end{center}
\vspace{-0.4cm}
\end{table*}

\subsection{Cross--validation}

We devise a cross--validation scheme so that each classification is trained on a manageably sized training set.
We first divide the entire data matrix ${\tt \tilde x_{ijk}}$
into ${\tt nk}=5$ subsets, each of which preserving the proportion among the different classes as ${\tt \tilde x_{ijk}}.$  
We keep one of the ${\tt nk}$ subsets as test set with the original class distribution and resample the remaining ${\tt nk}-1$ subsets, producing the resampled data. 
We then divide the resampled data into ${\tt nk^{\prime}}=3$ subsets, one of which serving as the resampled input data. 
We then divide the resampled input data into ${\tt nk^{\prime\prime}}=5$ subsets, one of which serving as resampled test set and the remaining serving as resampled training set. 
We rotate the resampled training data set over the ${\tt nk^{\prime\prime}}$ subsets so that the fitting of the classification model is done ${\tt nk^{\prime\prime}}$ times on different training sets.
We then rotate the resampled input data over the ${\tt nk^{\prime}}$ sets so that the fitting of the classification model is done ${\tt nk^{\prime\prime}}\times{\tt nk^{\prime}}$ times on different training sets. 

We average the ${\tt nk^{\prime\prime}}\times {\tt nk^{\prime}}$ classification predictions over ${\tt nk^{\prime\prime}},$ yielding ${\tt nk^{\prime}}$ mean predictions of the test data. 
We then average the ${\tt nk^{\prime}}$ classification predictions, yielding one confusion matrix labeled by the corresponding ${\tt nk^{\prime}}$ mean accuracies. 
 
The fitting resulting from each of the 
${\tt nk^{\prime\prime}}\times {\tt nk^{\prime}}$ 
subsets is applied to the test data, thus producing 
${\tt nk^{\prime\prime}}\times {\tt nk^{\prime}}$ classification predictions for each element in the test data. The final combined prediction is the mean over the 
 ${\tt nk^{\prime\prime}}\times {\tt nk^{\prime}}$ classification predictions, whose resulting accuracies we include in the tables.

\subsection{Selection of neural network architecture}

In order to select an adequate NN architecture, we try different architectures for RNN and CNN \cite{brownlee_arch}. 
We use the 
implementation from {\sc TensorFlow}\footnote{{\url https://www.tensorflow.org}} via the application programming interface {\sc Keras}\footnote{{\url https://keras.io}}.  

We represent the NN architectures schematically as sequences of layers,  
with Input representing the input data and Output representing the 
predicted classification. We explore four types of architectures: 
\paragraph{architecture with LSTM layers, named ${\tt LSTM}$}
\ba
{\tt LSTM:}~
\text{Input} \to \text{LSTM} \to \text{Dropout} \to \text{Dense} \to \text{Output};
\ea
\paragraph{architecture with 
Conv1D layers, named ${\tt Conv}$}
\ba
{\tt Conv:}~
\text{Input} &\to& \text{Conv1D} \to \text{Dropout} \to \text{MaxPool} \cr
&\to& \text{Flat} \to \text{Dense} \to \text{Output};
\ea
\paragraph{arquitectures that combine both LSTM and Conv1D layers, named ${\tt Conv+LSTM}$ and ${\tt ConvLSTM}$}
\ba
{\tt Conv+LSTM:}~
\text{Input} \!\!\!&\to & \!\!\!\text{Conv1D} \to \text{Dropout} \to \text{MaxPool} \cr
&\to& \!\!\!\text{Flat} \to \text{Dense}  \to  \text{LSTM} \to \text{Dropout}\cr
&\to& \!\!\!\text{Dense} \to \text{Output},\\ 
\cr
{\tt ConvLSTM:}~ 
\text{Input} &\to& \text{ConvLSTM} \to \text{Dropout} \cr
&\to& \text{Flat} \to \text{Dense} \to \text{Output}.
\ea

We first test these architectures for the minimal NN formulation and for approximately the same number of layers, setting the kernel size to ${\tt kernel\_size}=4$ and the number of filters to ${\tt n\_filter}=64.$ 
As measures of performance, we use the total number of true positives (TP) and the number of predicted classes; as measure of efficiency, we use the running time (Table ~\ref{table:test_arch}).

We observe that the ${\tt Conv}$ architecture yields the best performance 
and the shortest running time;  
this prompted us to consider ${\tt Conv}$ for further study. 
We also observe that most heart beats belonging to the classes $\{{\tt /, A, L, N, R, a, e, j}\}$ are correctly classified by all architectures. The goal is now 
to increase the number of TP of the other classes, namely $\{{\tt F, J, V, f}\}.$ 

\begin{figure*}
\centerline{
\includegraphics[width=7.cm]{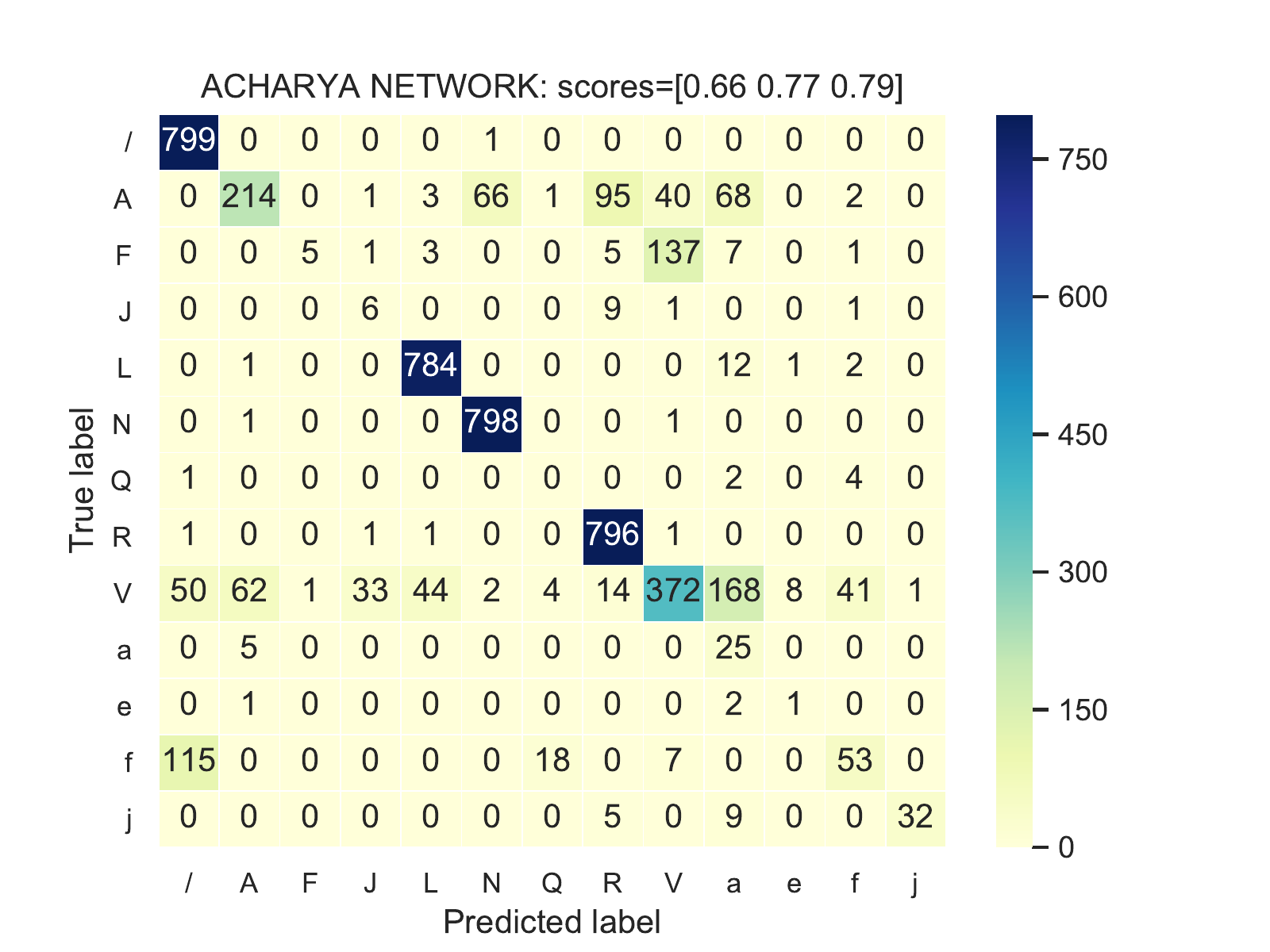}
\hspace{-1.5cm}
\includegraphics[width=7.cm]{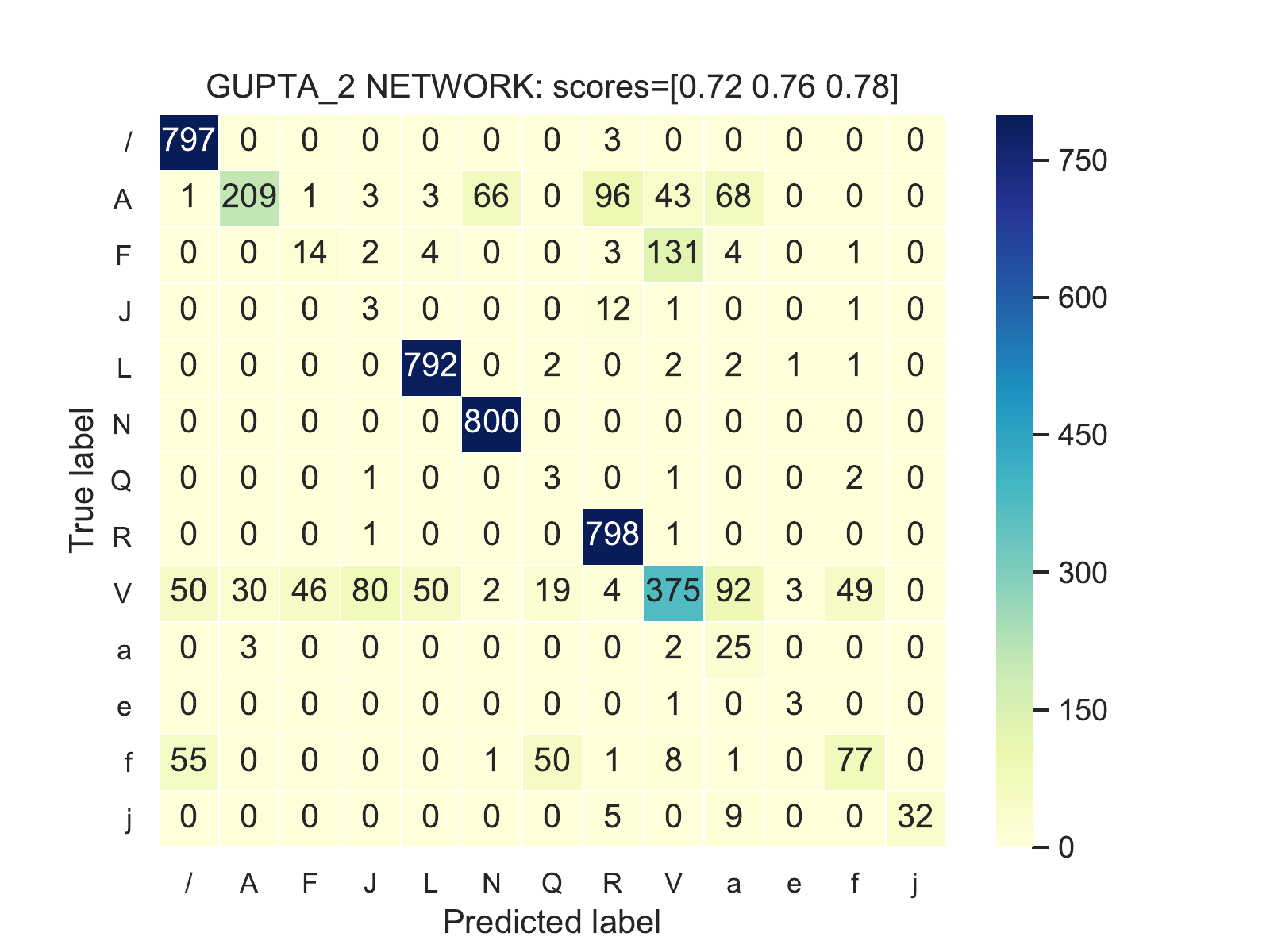}
\hspace{-1.5cm}
\includegraphics[width=7.cm]{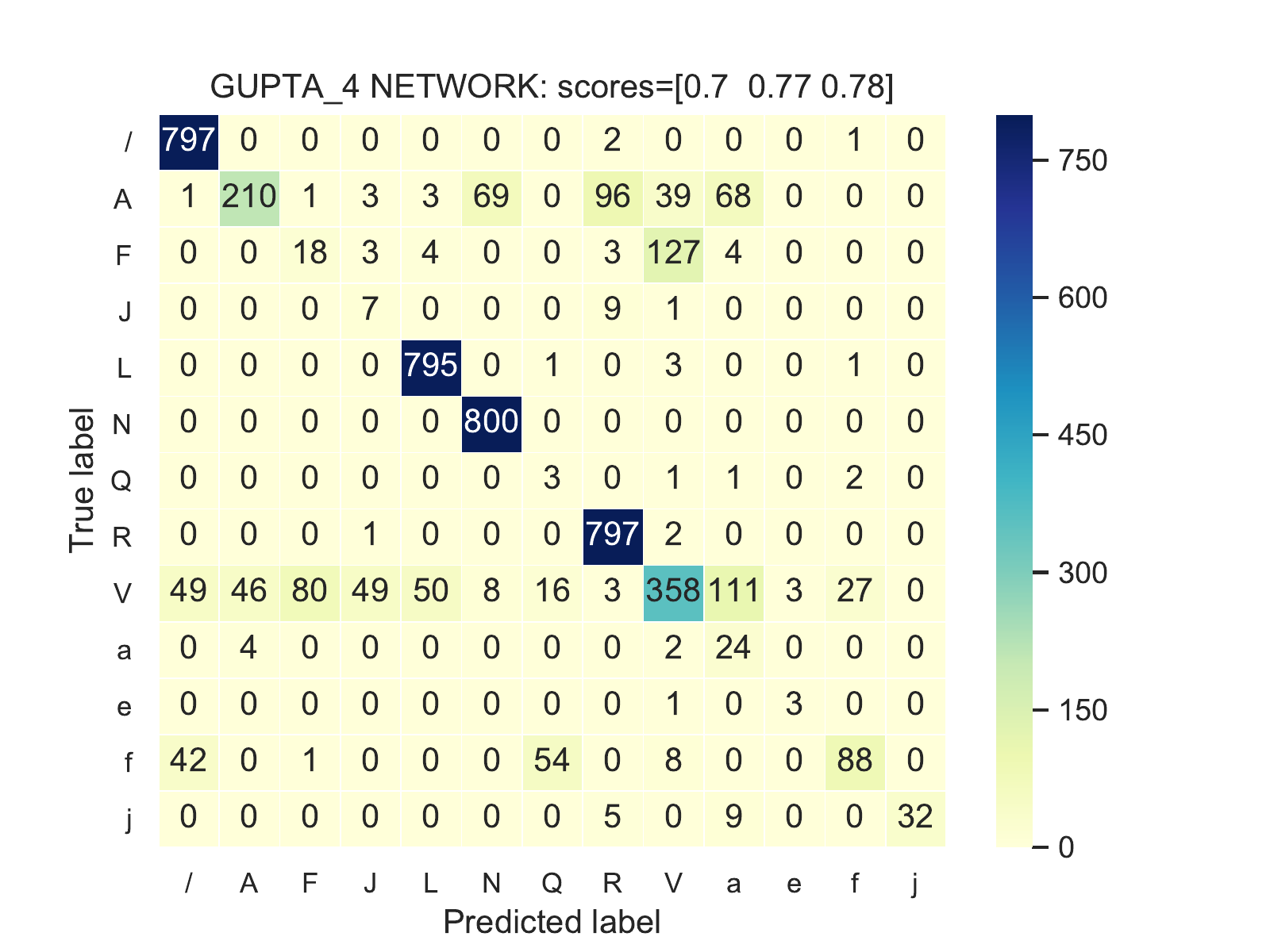}
}
\vspace{-0.2cm}
\caption{\baselineskip=0.4cm{
{\bf Confusion matrix from  NNs. } 
Left panel: ${\tt Acharya}$ with ${\tt n\_conv}=3$ convolutional layers, ${\tt kernel\_size}=\{4, 4, 4\},$ ${\tt n\_pool}=3$ MaxPool layers and ${\tt n\_drop}=3$ drop--out layers. 
Centre panel: ${\tt Gupta\_2}$ with ${\tt n\_conv}=6$ convolutional layers, ${\tt kernel\_size}=\{4,(4,4), (4,4,4)\},$ ${\tt n\_pool}=3$ AveragePool layers and ${\tt n\_drop}=3$ drop--out layers.
Right panel: ${\tt Gupta\_4}$ with ${\tt n\_conv}=7$ convolutional layers, ${\tt kernel\_size}=\{4,(4,4), (4,4,4), 4\},$ ${\tt n\_pool}=4$ AveragePool layers and ${\tt n\_drop}=3$ drop--out layers.
All NNs have ${\tt padding}={\tt valid}.$ 
}}
\label{fig:cm_new}
\vspace{-0.2cm}
\end{figure*}

\subsection{Selection of convolutional neural network architecture}

We 
test the ${\tt Conv}$ architecture for different number and organisation of layers. We explore eight NNs.  
In the Conv1D layers, we first set ${\tt kernel\_size}=4$ and ${\tt padding}={\tt valid},$ and vary the number of filters within the range ${\tt n\_filter}\in \{16, 32, 64\},$ starting at 16 and doubling every time that the Conv1D layer is preceded by a pooling layer. 
As measure of performance, we use the total TP. 

We start with the NN suggested in Ref.~\cite{acharya_2017}, 
since this NN was conceived to classify heart beats from the same database. We keep the architecture as shown below, named ${\tt Acharya:}$
\ba
{\tt Acharya:}~ 
\text{Input} \!\!\!\!& \to & \!\!\!\!\text{Conv1D (ReLU)} \to \text{Dropout} \to \text{MaxPool} \cr
&\to & \!\!\!\!\text{Conv1D (ReLU)} \to \text{Dropout} \to \text{MaxPool} \cr
&\to & \!\!\!\!\text{Conv1D (ReLU)} \to \text{Dropout} \to \text{MaxPool} \cr
&\to & \!\!\!\!\text{Flat} \cr 
&\to & \!\!\!\!\text{Dense (ReLU)} \to \text{Dense (ReLU)} \cr 
&\to & \!\!\!\!\text{Dense (Softmax)} \to \text{Output}.
\ea

Comparing 
${\tt Acharya}$ (Fig.~\ref{fig:cm_new}, left panel) with 
the previous best performing NN,  
the TP of 
$\{\tt f\}$ increases significantly, whereas the TP of 
$\{\tt F, V\}$ decrease, with the total TP staying approximately the same. 

We 
change ${\tt Acharya}$ by moving the drop--out layers from after the Conv1D layers to after the dense layers, which we name ${\tt Acharya\_2}$. 
Comparing ${\tt Acharya\_2}$ 
with ${\tt Acharya},$ 
the TP of 
$\{\tt F\}$ increases, whereas the TP of 
$\{\tt V, f\}$ decrease, with the total TP decreasing, thus a worsening in performance. 

We  
test the NN suggested in Ref.~\cite{gupta_2018}, 
since this NN was conceived to estimate cosmological parameters, which requires looking for different scales in the data. We keep the architecture as shown below, named ${\tt Gupta:}$
\ba
{\tt Gupta:}~
\text{Input} & \to & \text{Conv1D (ReLU)} \to \text{AveragePool} \cr
&\to & \text{Conv1D (ReLU)} \cr
&\to & \text{Conv1D (ReLU)} \to \text{AveragePool} \cr
&\to & \text{Conv1D} \to \text{AveragePool} \cr
&\to & \text{Conv1D} \to \text{AveragePool} \cr 
&\to & \text{AveragePool} \to \text{Flat} \cr
&\to &\text{Dense (ReLU)} \to \text{Dropout} \cr 
&\to &\text{Dense (ReLU)} \to \text{Dropout} \cr 
&\to &\text{Dense (ReLU)} \to \text{Dropout} \cr 
&\to& \text{Dense (Softmax)} \to \text{Output}.
\ea
This NN contains contiguous convolutional layers without intermediate pooling layers, 
forming a block of two Conv1D layers. 
Comparing 
${\tt Gupta}$  
with  
${\tt Acharya},$ 
the TP of  
$\{{\tt F, f}\}$ increase but the TP of 
$\{{\tt V}\}$ decreases, with the total TP decreasing, 
thus no improvement in performance.

We 
change ${\tt Gupta}$ by adding 
another set of contiguous Conv1D layers without intermediate pooling layers, as shown below, named ${\tt Gupta\_2:}$
\ba
{\tt Gupta\_2:}~
\text{Input} & \to & \text{Conv1D (ReLU)} \to \text{AveragePool} \cr
&\to & \text{Conv1D (ReLU)} \cr
&\to & \text{Conv1D (ReLU)} \to \text{AveragePool} \cr
&\to & \text{Conv1D (ReLU)} \cr
&\to & \text{Conv1D (ReLU)} \cr
&\to & \text{Conv1D (ReLU)} \to \text{AveragePool} \cr
&\to & \text{Flat} \cr
&\to &\text{Dense (ReLU)} \to \text{Dropout} \cr 
&\to &\text{Dense (ReLU)} \to \text{Dropout} \cr 
&\to &\text{Dense (ReLU)} \to \text{Dropout} \cr 
&\to& \text{Dense (Softmax)} \to \text{Output}
\ea 
Comparing 
${\tt Gupta\_2}$ (Fig.~\ref{fig:cm_new}, centre panel) with 
${\tt Gupta},$ 
the TP of  
$\{{\tt V}\}$ increases but the TP of $\{{\tt F, f}\}$ decrease, with the total TP increasing, thus an improvement in performance.

We change ${\tt Gupta\_2}$ by adding drop--out layers before each pooling layer in a similar way to ${\tt Acharya},$ which we name  ${\tt Acharya\_3.}$ 
Comparing  
${\tt Acharya\_3}$ with 
${\tt Gupta\_2},$ 
the total TP stays approximately the same, 
thus no improvement in performance.

We change again ${\tt Gupta\_2}$ by adding batch--normalization layers (BatchNorm) between the Conv1D layers and the ReLU layers, 
which we name ${\tt Gupta\_3.}$ Comparing  
${\tt Gupta\_3}$  
with 
${\tt Gupta\_2},$  
the total TP stays approximately the same, 
thus no improvement in performance. 

While the newly generated Gupta NNs prove to increase the TP of 
$\{{\tt V} \}$ in comparison to  
${\tt Gupta,}$ they can not increase the TP of  
$\{{\tt F, J, f}\}.$
Hence we change ${\tt Gupta\_2}$ again by adding a single Conv1D layer just before the flat layer, 
which we name ${\tt Gupta\_4.}$
Comparing  
${\tt Gupta\_4}$ (Fig.~\ref{fig:cm_new}, right panel) 
with 
${\tt Gupta\_2}$ 
or ${\tt Gupta\_3,}$  
the total TP decreases slightly, thus a worsening in performance. 

We change ${\tt Gupta\_4}$ by adding BatchNorm layers between the Conv1D layers and the ReLU layers, 
which we name ${\tt Gupta\_5.}$ Comparing  
${\tt Gupta\_5}$ 
with  
${\tt Gupta\_2}$ 
or ${\tt Gupta\_3},$  
the total TP decreases slightly, thus a worsening in performance. 

We thus select ${\tt Gupta\_2}$ for 
further testing. 

\begin{table*}[t]
\caption{\baselineskip=0.4cm{
{\bf Comparison with previously published work.} 
Column 1: Identification of the work. 
Column 2: Identification of the network size. 
Column 3: Identification of the number of classes in the data. 
Columns 4-6: Identification of the performance metrics for the best performing network, 
where Accuracy is the resulting accuracy, and $\left<\text{Precision}\right>_{w}$ and $\left<\text{Recall}\right>_{w}$ are respectively the mean precision and mean recall over all classes, weighted by the size of each class.
}}
\label{table:compare_nn}
\begin{center}
\vspace{-0.2cm}
\begin{tabular}{c|cc|ccc}
Work 
& Network size & No. classes  
& Accuracy & $\left<\text{Precision}\right>_{w}$ & $\left<\text{Recall}\right>_{w}$  \\
\hline
Acharya et al. (2017) \cite{acharya_2017}	
& 3 Conv1D + 3 Dense & 5& (0.935, 0.940) 
& (0.979,  0.979) & (0.960, 0.967)\\
He et al. (2018) \cite{he_2018}	
& 9 Conv1D + 2 Dense & 5 & (0.979, 0.988) 
& &\\
Jun et al. (2018) \cite{jun_2018}		
& 6 Conv2D + 1 Dense 
& 8 & 0.990 & 0.986 & 0.978\\
Rajpurkar et al. (2017) \cite{rajpurkar_2017} 
& 33 Conv1D + 1 Dense 
& 14 & & 0.809 & 0.827\\
Carvalho (2020) 		
& 6 Conv1D + 4 Dense & 13  & 0.821 & 0.848 & 0.822\\
\end{tabular}
\end{center}
\vspace{-0.2cm}
\end{table*}

\subsection{Selection of convolutional neural network hyper--parameters}

We 
explore further the ${\tt Gupta\_2}$ NN by varying some hyper--parameters. 
As measures of performance, we use the average accuracies, the accuracy of the 
resulting accuracy and the total TP.

\paragraph{Type of pooling layer and type of padding}
We test the type of pooling layer and the type of padding simultaneously. For each type of pooling layer, we vary the type of padding in the Conv1D layers, while keeping the kernel sizes per block equal to ${\tt kernel\_size}=\{4,(4,4), (4,4,4)\}.$ 
We observe that AveragePool with ${\tt padding}={\tt same}$ yield the best performance. 

\paragraph{Number of dense layers}
We 
vary the number of dense layers within the range ${\tt n\_dense}\in \{1, 2, 3, 4, 5, 6\},$ where the number of drop--out layers is ${\tt n\_drop}={\tt n\_dense}-1,$ while keeping
the number of filters equal to ${\tt n\_filter}=64.$ 
We observe that ${\tt Gupta\_2}$ yields an increase in the total TP up to ${\tt n\_dense}=4$ and a decrease for ${\tt n\_dense}>5.$ Since the difference in performance between ${\tt n\_dense}=4$ and ${\tt n\_dense}=5$ is not significant, 
we keep the original ${\tt Gupta\_2}$ with ${\tt n\_dense}=4$ dense layers due to its simplicity.

\paragraph{Kernel sizes of each convolutional layer}
We 
vary the kernel sizes 
of each block over the range $\{4, 8, 16\},$ arranging these three values in combinations respectively of one, two and three values in consecutive order. 
With these constraints, the best performing NNs of the ${\tt Gupta\_2}$ type are for 
\ba
{\tt kernel\_size} \in  \!\!\!\!\!\!
&\big\{ & \!\!\!\!\!\!\! \{4, (4,4), (4,4,4)\}, \!\{4, (8,8), (4,8,16)\},\cr
&& \!\!\!\!\!\!\!\! \{8, (4,4), (4,4,4)\},  \!\{8, (16,16), (4,4,4)\}, \cr
&& \!\!\!\!\!\!\!\! \{16, (8,4), (4,8,16)\},\!\{16, (8,8), (4,4,4)\} \!\big\}.~~~~
\ea 
While all these NNs yield average accuracies between 0.7 and 0.8, the NN with ${\tt kernel\_size}=\{16, (8,8), (4,4,4)\}$ yields the largest total TP and largest resulting accuracy.
This suggests that the important features for the classification of heart beats are best captured in groups of 4, 8 and 16 data points, with the convolutions  
proceeding from large to small scales. These scales correspond to $\{4, 8, 16\}~\times~{\tt dt}=\{0.011, 0.022, 0.044\}~\rm{s}.$ 
We thus select ${\tt kernel\_size}=\{16, (8,8), (4,4,4)\}$ for further testing. 

\subsection{Selection of input data}
We 
test  
${\tt Gupta\_2}$ NN 
for different data matrices,  
namely ${\tt \tilde x_{ijk}}=\{{\tt x_{ij,sign1}}, {\tt x_{ij,sign2}}\},$ consisting of the original variables,  
and  
${\tt \tilde x_{ijk}}=\{{\tt x_{ij,sign1}}, {\tt x_{ij,sign2}}, {\tt  \vert X_{ij,sign1} \vert }\},$ consisting of the variables selected from the correlation constraints. 
 
While 
both data matrices yield mean accuracies between 0.7 and 0.8, the 
data matrix 
${\tt \tilde x_{ijk}}=\{{\tt x_{ij,sign1}}, {\tt x_{ij,sign2}}\}$ without smoothing yields the largest total TP and the largest resulting accuracy. 
This suggests that 
the Fourier transforms of the original signals do not add discriminative information 
and that smoothing the heart beats might erase important features. 

\begin{figure}
\centerline{
\includegraphics[width=9.5cm]{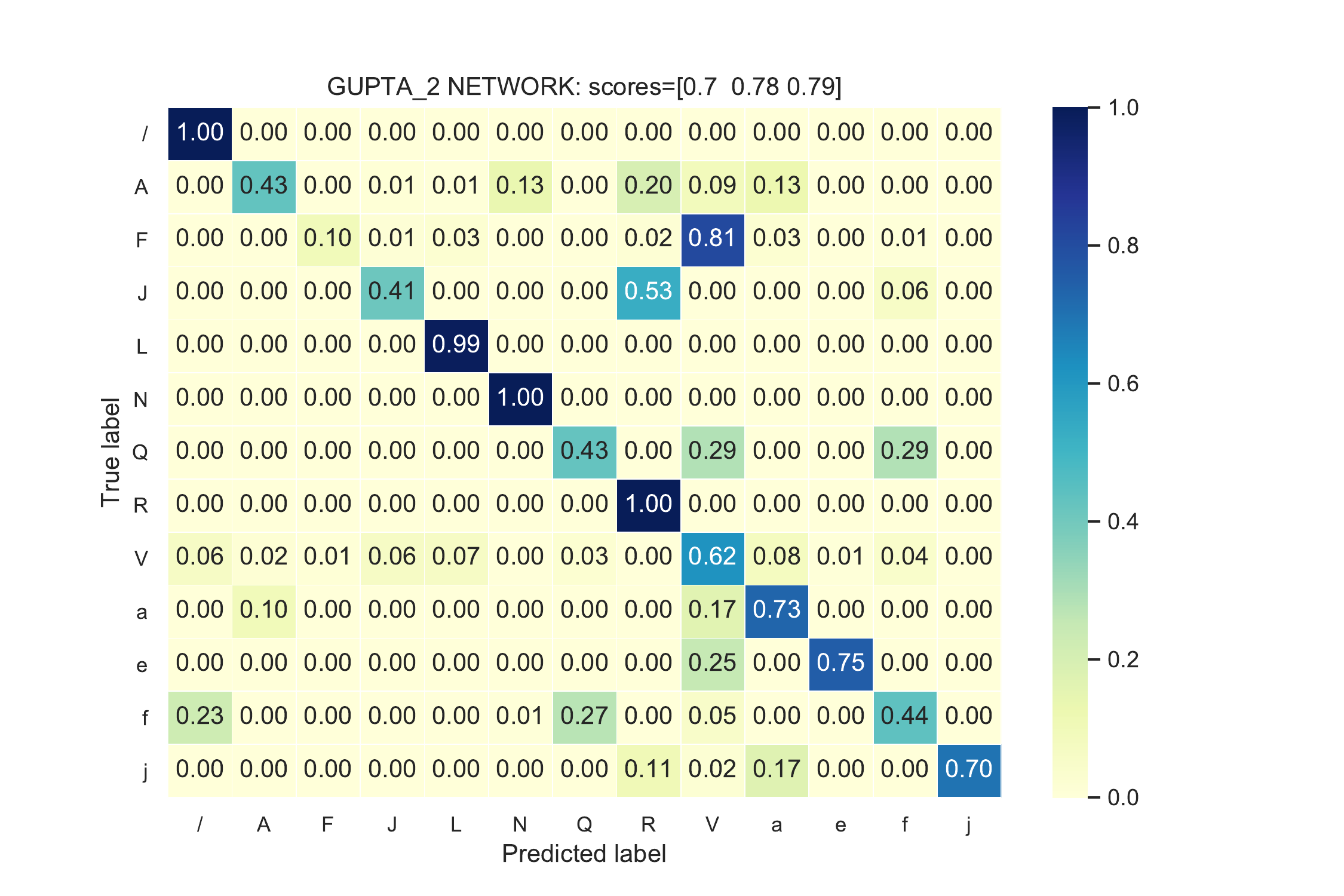}}
\vspace{-0.2cm}
\caption{\baselineskip=0.4cm{
{\bf Normalised confusion matrix from selected NN.}
Gupta\_2 NN with ${\tt kernel\_size}=\{16,(8,8), (4,4,4)\},$ AveragePool layers and ${\tt padding}={\tt same},$ 
applied to ${\tt \tilde x_{ijk}}=\{{\tt x_{ij,sign1}}, {\tt x_{ij,sign2}}\}.$
}}
\label{fig:cm_gupta_2_kernel_16_8_8_4_4_4}
\vspace{-0.2cm}
\end{figure}

\section{Results}
\label{sec:results}

\subsection{Results from the best--performing neural network}
We produce the distribution of the predicted heart beat classes, which
follows approximately the same distribution as that of the original heart beat classes (Fig.~\ref{fig:hist_class}, bottom panel).
We produce the confusion matrix of our best performing neural network 
normalised to the data per heart beat class for easier assessment of the performance per class (Fig.~\ref{fig:cm_gupta_2_kernel_16_8_8_4_4_4}). 
The class that is worst classified is $\{{\tt F}\}$ (``Fusion of ventricular and normal beat"), which is mostly misclassified as $\{{\tt V}\}$ (``Premature ventricular contraction"), hence the neural network is confounding between two ventricular arrhythmias. 
The next worse classified classes are:
a) $\{{\tt J}\}$ (``Nodal (junctional) premature beat"), which is mostly classified as $\{{\tt R}\}$ ( ``Right bundle branch block beat");
b) $\{{\tt A}\}$ (``Atrial premature beat"), which is also classified as either $\{{\tt N, R, V, a}\}$ (where {\tt a} stands for ``Aberrated atrial premature beat");
and c) $\{{\tt f}\}$ (``Fusion of paced and normal beat"), which is also classified as either $\{{\tt /, Q}\}$ (respectively ``Paced beat" and ``Unclassifiable beat"). 

\subsection{Comparison with results from other published work}
We compare our results with those of other NNs developed in previously published work to classify cardiac arrhythmias from annotated ECG records (Table ~\ref{table:compare_nn}). 

On the properties of the NNs, we note that most suggested neural networks have around 10 layers, with the exception of the neural network suggested in Ref.~\cite{rajpurkar_2017} which has over 30 layers. Such a size is not justified  
by the authors. 

On the number of arrhythmias, Refs.~\cite{acharya_2017, he_2018} used 
another annotation system consisting of five classes only.
References~\cite{rajpurkar_2017, jun_2018} used the MIT--BIH adopted annotation system. 

On the 
performance metrics, while Refs.~\cite{acharya_2017, jun_2018} used data resampling, 
it is not clear whether the reported results  
relate to a test subset of the original data or to the resampled data. 

In conclusion, our results are 
better than those obtained in Ref.~\cite{rajpurkar_2017} for about the same number of  
classes and  
less than a third of the number of layers, albeit worse than the results  
obtained for 
about half the number of 
classes with about the same number of layers.  
This attests to the competitiveness of our method and its adequateness for a solution deployment. 



\end{document}